\documentstyle[aps,prb,twocolumn,epsf]{revtex}

\begin{document}

\twocolumn[
\hsize\textwidth\columnwidth\hsize\csname@twocolumnfalse\endcsname

\title{Strong-coupling branching of FQHL edges}

\author{V.V. Ponomarenko and D.V. Averin}

\address{Department of Physics and Astronomy, Stony Brook University, 
SUNY, Stony Brook, NY 11794, USA}

\date{\today}

\maketitle

\begin{abstract}
We have developed a theory of quasiparticle backscattering 
in a system of point contacts formed between single-mode 
edges of several Fractional Quantum Hall Liquids (FQHLs) with 
in general different filling factors $\nu_j$ and one common 
single-mode edge $\nu_0$ of another FQHL. In the strong-tunneling 
limit, the model of quasiparticle backscattering is obtained   
by the duality transformation of the electron tunneling model. 
The new physics introduced by the multi-point-contact geometry of 
the system is coherent splitting of backscattered quasiparticles 
at the point contacts in the course of propagation along the 
common edge $\nu_0$. The ``branching ratios'' characterizing the 
splitting determine the charge and exchange statistics of the edge 
quasiparticles that can be different from those of Laughlin's 
quasiparticles in the bulk of FQHLs. Accounting for the edge 
statistics is essential for the system of more than one point 
contact and requires the proper description of the flux 
attachement to tunneling electrons. 
\end{abstract}



]

\section{Introduction}

Electron transport properties of the Fractional Quantum Hall Liquids 
(FQHLs) continue to attract considerable interest motivated by the 
unusual properties of FQHL excitations which are characterized 
by fractional charge and non-trivial exchange statistics. Since 
excitations in the bulk of a typical FQHL are suppressed by the 
energy gap, the low-energy transport properties are determined by 
the gapless modes at the edges of the liquid \cite{b1}. In the 
simplest case of a FQHL with fillings factor $\nu=1/odd$, its 
sharp edge should support one bosonic mode that is described as a 
single branch of chiral Luttinger liquid \cite{b4}. Although the 
edge states are gapless and in principle can carry excitations of 
arbitrary charge, in the case of tunneling between the edges of 
FQHLs with equal filling factors, strong tunneling produces 
excitations which coincide \cite{b19} with Laughlin's quasiparticles 
in the bulk of the liquid. Experiments on resonant tunneling through 
a quantum antidot \cite{b8} and on the shot noise in quasiparticle 
backscattering in a single point contact \cite{b9} measure directly 
the fractional charge of these excitations. Edge-state tunneling 
should also make possible the measurements of quasiparticle exchange 
statistics \cite{b21} which can be used in the development of FQHE 
qubits \cite{b24} for solid-state quantum computation. 

In the situation of tunneling between the edges of FQHLs with 
different filling factors, strong tunneling should produce 
quasiparticles that are different from bulk quasiparticles 
\cite{b17}. The charge of such ``contact'' 
quasiparticles coincides with the dc conductance of the point 
contact (if the two are measured in units of electron charge $e$ 
and the free-electron conductance $e^2/h$, respectively.) 
Theory up to now could predict only the properties of one point 
contact between different edges. The aim of this work is to 
develop a theory of strong tunneling between FQHL edges with 
different filling factors in a junction with more than one 
point contact. Such a model of multi-point-contact junction was 
introduced \cite{b11,b12} to describe experiments \cite{b26} on 
tunneling between FQHL edge and external Fermi-liquid reservoir. 
Understanding of the strong-tunneling limit of this model is 
important for the description of the reservoir-edge 
equilibration that is responsible for correct quantization of 
the two-terminal FQHL conductance.\cite{b13} 
Multi-point-contact junctions with a well-controlled geometry 
similar to the one considered in this work are also studied 
experimentally.\cite{b28} 

The main technical obstacle to the description of strong 
multi-point-contact tunneling between different edges is the fact 
that the bosonic fields in the tunneling operators of different 
contacts do not commute with each other and, at first sight, can 
not be localized simultaneously at the minima of large tunnel 
potential, as can be done for one point contact. This problem is 
resolved, however, if the bosonic fields are modified\cite{b13} 
by the proper choice of the statistical phase of the tunneling 
electrons [see Eqs.~(\ref{e11}) and (\ref{e12}) below]. The 
statistical phase preserves the Fermi statistics of electrons 
but accounts for the change of the number of flux quanta 
attached to them in the FQHLs \cite{b29} in the process of 
tunneling between the liquids with different filling factors. 
With the flux attachement taken into account, the total 
bosonic tunneling fields commute, and one can build the 
strong-coupling description of the multi-point-contact junction 
in close analogy to the case of one point contact. Here, we use 
this approach to develop complete description of strong 
tunneling in the multi-point-contact junction extending previous 
results \cite{b13} to quasiparticle backscattering. 

Important elements of our approach can be summarized as 
follows. Specific junction model considered in this work is 
characterized by the existence of one edge $\nu_0$ common to 
all point contacts (Fig.~1). The quasiparticle backscattering 
is produced by finite reflection coefficients of the contacts, 
and is described as instanton tunneling between the infinite set 
of the ground states of the original electron tunneling model 
that are degenerate in the absence of backscattering. 
Expansion of the junction dynamics in terms of instantons gives 
the model of quasiparticle tunneling which is dual to the 
electron tunneling model. Duality transformation relating the 
two models produces the quasiparticle exchange statistics from 
the electron statistical phases ascribed through the flux 
attachement. Correct description of quasiparticle statistics 
in our junction geometry enables one to combine sequentially 
the edge-state transformations at different point contacts in 
such a way that the incoming edges are split coherently at each 
point contact in the course of propagation along the common 
edge $\nu_0$. This process can be described with the ``branching 
ratios'' which define how the incoming currents (or charges) are 
split between the outgoing edges. Quasiparticles produced by 
backscattering at one point contact are split similarly at all 
point contacts downstream the edge $\nu_0$ from the ``contact 
of origin'', and the edge-state branching ratios determine 
the charges generated in each edge by backscattering. As usual, 
these charges should manifest themselves through the intensity 
of the shot noise of outgoing currents. The charge splitting 
also provides natural interpretation of the quasiparticle 
exchange statistics obtained by the duality transformation, so 
that both the charge and statistics of backscattered 
quasiparticles can be viewed as being defined by the 
strong-coupling edge-state branching. 

The paper is organized as follows. Section II defines the electron 
tunneling model considered in this work. Section III describes 
the bosonization procedure for the Klein factors of electron 
tunneling operators that implements the flux attachment. Section 
IV explains the instanton transformation, formulates the dual 
model of quasiparticle tunneling, and calculates the dc current 
and shot noise in outgoing edges of the junction. In Section V
we apply the results of Sec. IV to the situation that corresponds 
to the junction between a Fermi-liquid reservoir and FQHL edge.  
Quasiparticle backscattering in this case gives rise to corrections 
to the Ohmic behavior of the junction: non-linear current-voltage 
characteristic and shot noise, which limit the reservoir-edge 
equilibration obtained in the absence of backscattering. In the 
Appendix, we summarize for convenience the known results for 
strong tunneling in one point contact that are used in the main 
text.

\section{Multi-edge model of weak electron tunneling}

The model we consider consists of $n$ edges of different FQHLs
of the filling factors $\nu_j$, $j=1 \, ...\, n$, which form
$n$ tunnel point contacts with an edge of another FQHL
of the filling factor $\nu_0$ (see Fig.\ 1). All liquids are
assumed to correspond to simple Laughlin states with the filling
factors $\nu_l=1/odd$, $l=0\, ...\,n$, which in general can all
be different. Since bulk excitations inside the liquids have
energy gap, at energies below some common cut-off energy $D$, only
the edges support excitations that propagate along them. In the
weak tunneling regime, when each of the $j$th edges is well
separated from the $0$ edge, the tunnel currents between them
should be carried by individual electrons. We assume that the
tunnel contact formed by the $j$th edge is located at the point
$x_j$ along the $0$ edge and its width is much smaller than the
magnetic length. Tunneling in such a system of contacts is
described by the Lagrangian:
\begin{equation}
{\cal L}_{tunn}= \sum_{j=1}^{n} [U_j\psi_0^+(x_j,t)
\psi_j(x_j,t)+ h.c.]\, ,
\label{e1} \end{equation}
where the tunnel amplitudes $U_j$ are taken to be real and
positive, and $\psi_l$ is the electron operator of the
$l$th edge. 

\begin{figure}[htb]
\setlength{\unitlength}{1.0in}
\begin{picture}(3.,1.1)
\put(.05,.1){\epsfxsize=3.0in\epsfbox{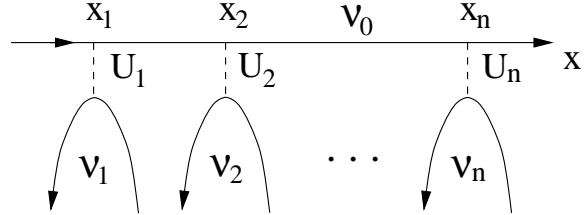}}
\end{picture}
\caption{Diagram of the model of a tunnel junction considered
in this work: $n$ edges of FQHLs with in general different
filing factors $\nu_j$ tunnel into one FQHL edge with the filing
factor $\nu_0$ at points $x_j$ along this edge. $U_j$ are
electron tunneling amplitudes. Each edge is assumed to support
one branch of bosonic excitations with arrows indicating 
direction of propagation of these excitations. }
\end{figure}

If we assume that all edges are sharp enough to enable us 
to use simple bosonic single-mode description of the edge-state 
dynamics introduced by Wen \cite{b4}, the electron operator 
$\psi_l$ can be expressed as
\begin{equation}
\psi_l=(2\pi \alpha)^{-1/2} \xi_l e^{i\phi_l(x,t)/
\sqrt{\nu_l}} \, .
\label{e2} \end{equation}
Here $\phi_l$ is the bosonic field of the edge $l$; Majorana 
fermions $\xi_l$ account for mutual statistics of electrons in 
different edges, and a common factor $1/\alpha=D/v$ denotes 
momentum cut-off of the edge excitations. Since the spatial 
dynamics of the edges $j=1\, ...\,n$ does not affect the tunnel 
currents, propagation velocities of these edges are irrelevant, 
and we take them to be equal to the velocity $v$ of the 
excitations of the edge $\nu_0$. The bosonic fields $\phi_l$ 
are normalized in such a way that their free dynamics is 
governed by the standard Lagrangian density \cite{b4}
\begin{equation}
{\cal L}_0 =\frac{1}{2} \sum_{l,p=0}^n \phi_l
\hat{G}^{-1}_{lp} \phi_p \equiv \frac{1}{2}
\phi \hat{G}^{-1} \phi \, ,
\label{e3} \end{equation} 

\vspace{-2ex} 

\[ \hat{G}^{-1}_{lp} =\frac{\delta_{lp}}{ 2 \pi} \partial_x
(\partial_t+ v\partial_x) \, , \]
which corresponds to the Hamiltonian
\begin{equation}
{\cal H}_0=\frac{v}{4 \pi} \int dx (\partial_x \phi_l (x))^2
\label{e6} \end{equation}
and the commutation relations
\begin{equation}
[\phi_l(x),\phi_p(0)]=i \pi \,
\delta_{lp}\mbox{sgn}(x) \, . \label{e7} \end{equation}

In the following, we will need free correlator of the
fields $\phi_l$ ordered in imaginary time $\tau$:
\begin{equation}
\hat{G}_{lp}(x,\tau)=\delta_{lp}g(x,\tau)\, , \;\;\;
g(x,\tau)= \langle T_{\tau} \{ \phi_l(x,\tau)
\phi_l(0,0) \} \rangle \, . \label{e4} \end{equation} 
To find this correlator, we start with the retarded Green's 
function in real time $t$:
\begin{equation}
g^R(x,t)=-i\Theta(t) \langle [\phi_l(x,t), \phi_l(0,0)] 
\rangle \, , \label{e5} \end{equation}
where $\Theta(t)$ is the step function. Lagrangian (\ref{e3})
implies that dynamics of $g^R(x,t)$ is governed by the
equation:
\[ \partial_x (\partial_t+ v \partial_x) g^R(x,t)= 2\pi
\delta (x) \delta (t) \, . \]
Solution of this equation that satisfies the symmetry of the
Eq.~(\ref{e5}) (the commutator part of Eq.~(\ref{e5}) should be
antisymmetric with respect to simultaneous change of sign of
$x,t$) is:
\[ g^R(x,t)=\pi \Theta(t) \mbox{sgn}(x-vt) \, . \]
The standard steps: Fourier transformation of $g^R(x,t)$, 
$g^R(x,\omega)=\int dt e^{i\omega t} g^R(x,t)$, and analytical 
continuation to the upper half-plane of frequency, $\omega 
\rightarrow i \omega$, give the Fourier transform of the 
imaginary-time correlator for positive frequencies $\omega$:
$g(x,\omega)= -g^R(x,\omega)|_{\omega \rightarrow i \omega}$. 

Similar calculations for the advanced Green's function 
\[ g^A(x,t)=i\Theta(-t) \langle [\phi_l(x,t), \phi_l]
\rangle = - \pi \Theta(-t) \mbox{sgn}(x-vt)
\, ,\]
give $g(x,\omega)$ for negative frequencies, and it can 
finally be written for all Matsubara frequencies as
\begin{equation}
g(x,\omega)=\frac{2 \pi}{\omega} \mbox{sgn} (x) \left(-
\frac{1}{2} +\Theta(\omega x)e^{-\omega x/v} \right) \,
. \label{e8} \end{equation} The first part of this 
correlator does not depend on $|x|$ and characterizes singular 
behavior of the correlator at short times, $g(x,\tau)\sim i 
\pi \mbox{sgn}(x\tau)/2$, which is responsible for the 
intra-edge statistics of the electronic operators $\psi_l$. 
With the adopted normalization of $\phi_l$, the operator of 
charge density of the $l$th edge at point $x$ is
\[ \rho_l(x,\tau)= (\sqrt{\nu_l}/2 \pi) \partial_x \phi_l(x,
\tau)\, , \]
and the current in the edge is related to the charge density
as $j_l=v \rho_l$.

As in the case of a single point-contact, the effective
electron tunneling amplitudes are renormalized from their 
original values $U_j$ in Eq. (\ref{e1}) by fluctuations of 
the fields $\phi_l$. The effective tunneling amplitudes scale 
with energy in such a way that the weak-tunneling limit is 
always stable at sufficiently low energy. The main focus of 
this work is on the strong-tunneling regime realized at large
temperatures or bias voltages $V_j$ across the point contacts 
which make the effective tunneling amplitudes large. The 
description of the strong-tunneling limit presented below is 
based on the instanton expansion around the saddle-point 
solution \cite{b13}. As discussed in the Introduction, this 
solution requires appropriate bosonization of the Klein 
factors of the electron tunneling operators that takes into 
account the flux attachement. 

\section{Bosonization of the Klein factors}

Substitution of the bosonic representation (\ref{e2}) of the
electron operators $\psi_l$ into Eq.~(\ref{e1}) transforms the
tunneling Lagrangian into
\begin{equation}
{\cal L}_{tunn} = \sum_{j=1}^{n} [{U_j\over 2 \pi \alpha} F_j
\exp \left\{i \lambda_j \varphi_j(t)\right\}+ h.c.]\, .
\label{e10} \end{equation}
Here
\[ \lambda_j \varphi_j (t) \equiv {\phi_0(x_j,t) \over
\sqrt{\nu_0}} - {\phi_j(x_j,t) \over \sqrt{\nu_j} } \, , \]
and the factors
\begin{equation}
\lambda_j=\left[ {\nu_0 +\nu_j\over \nu_0 \nu_j }\right]^{1/2}
\label{e9} \end{equation}
are chosen in such a way that the normalization of the bosonic
operator $\varphi_j$ coincides with the normalization of the
bosonic fields $\phi_l$ used before, so that the imaginary-time
correlator of $\varphi_j$ is given by the same Eq.~(\ref{e8})
with $x=0$: $g(0,\omega)=\pi/|\omega|$.

In the Lagrangian (\ref{e10}), $F_j$ represents the 
anticommuting statistical Klein factors, $F_j=\xi_0 \xi_j$, 
which account for mutual statistics of electrons in different 
edges. These factors play an important role in the strong 
multi-edge tunneling considered in this work. A convenient way 
of taking them into account is to express them through the 
zero-energy bosonic fields $\eta_j$:\cite{b13,b14} 
\begin{equation}
F_j=e^{i  \eta_j}, \;\;\; [\eta_i,\eta_j]=i\pi \gamma_{ij} \, ,
\label{e11} \end{equation}
where $\gamma_{ij}$ are odd integers. Different choices of
$\gamma_{ij}$ correspond to different branches of the phase of 
the Fermionic statistical factor $-1$ arising from interchange 
of $F_i$ and $F_j$. Under the condition that the point contacts 
are well-separated, $|x_i-x_j| \gg \alpha$, minimization of 
energy of the strong-coupling ground state requires that the 
statistical phases are taken as:\cite{b13}
\begin{equation}
\gamma_{ij} =\mbox{sgn}(i-j)(1- \delta_{ij}) \frac{1}{\nu_0} \, ,
\label{e12} \end{equation}
This equation assumes that the point contacts are numbered as
in Fig.\ 1: $x_i<x_j$ for $1 \le i<j \le n$, and the coordinate
$x$ increases in the direction of the edge propagation.

The steps similar to those that lead to Eq.~(\ref{e8})
give then the Fourier transform of the $\tau$-ordered
correlator of the fields $\eta_j$:\cite{rem}
\begin{equation}
\langle \eta_i \eta_j\rangle =\mbox{sgn}(i-j)(1- \delta_{ij})
\frac{\pi}{\omega \nu_0} \, .
\label{e13} \end{equation}
Substitution of the bosonized form (\ref{e11}) of
the Klein factors into Eq.~(\ref{e10}) turns the tunneling
Lagrangian into a function of $n$ bosonic variables
\begin{equation}
\Phi_j=\lambda_j \varphi_j+\eta_j \, ,
\label{e14} \end{equation}
and one can reduce the kinetic Lagrangian (\ref{e3}) to a
simplified form by integrating out all the fields except
$\Phi_j$. The resulting Gaussian action for the fields $\Phi_j$
can be obtained directly by noticing that it should be
determined by the matrix of correlators $\hat{K}_{ij} (\omega)=
\langle \Phi_i(-\omega) \Phi_j(\omega) \rangle $ as
\begin{equation}
{\cal S}_0=\frac{1}{2}\Phi \hat{K}^{-1} \Phi = \frac{1}{2}
\sum_{ij} \sum_{\omega} \Phi_i(-\omega)\hat{K}^{-1}_{ij}
(\omega) \Phi_j(\omega)\, ,
\label{e15} \end{equation}
where $\omega=2\pi mT$ denotes the Matsubara frequencies with
 $m=0 \pm 1, \pm2, ...$ and temperature $T$. The matrix of
correlators $\hat{K}_{ij}$ can be found from the correlator
(\ref{e8}) of the tunneling fields $\varphi_j$ (determined by
the original kinetic Lagrangian (\ref{e3})) and the correlators
(\ref{e13}) of the statistical fields. Combining the two we get
\begin{equation}
\hat{K}_{ij}(\omega)={2 \pi \over |\omega|}\left[{\lambda_j^2
\over 2}\delta_{ij}+ {\Theta( \mbox{sgn(i-j)}\omega) \over
\nu_0} e^{-|\omega t_{ij}|}(1- \delta_{ij}) \right] .
\label{e16} \end{equation}
where $t_{ij}=t_i-t_j$ and $t_j\equiv x_j/v$. Important feature 
of the terms in Eq.~(\ref{e16}) that are non-diagonal in $i,j$ 
is that they do not contain the singular statistical parts. The 
singular part of the correlator (\ref{e8}) of the tunneling 
fields $\varphi_j$ is cancelled by the statistical fields 
$\eta_j$, if the phases $\gamma_{ij}$ (\ref{e11}) are chosen 
appropriately, as in Eq.~(\ref{e12}).

\section{Strong-tunneling limit}

\subsection{Instanton expansion and duality transformation}

The tunneling action (\ref{e10}) expressed through the bosonic 
fields $\Phi_j$ (\ref{e14}) has an infinite set of minima at 
$\Phi_j=2 \pi \times \mbox{integer}$, $j=1\div n$, that are 
degenerate in energy. In the strong-tunneling limit, when the 
electron tunneling amplitudes $U_j$ are large, these minima are
well-separated in a sense that the amplitude of 
$\Phi_j$-tunneling between them is small. Such rare tunneling 
processes correspond physically to finite backscattering at each 
of the point contacts $j$ and are described by the instanton 
tunneling solutions which on the long-time scale behave as 
$2 \pi e_{lj} \Theta(\tau-\tau_{lj})$, where index $l$ counts 
different instanton tunneling events, $e_{lj}=\pm 1$, and 
$\tau_{lj}$ are the times of tunneling of the $\Phi_j$ component. 
On the short time scales on the order of $1/D$, the variation of 
$\Phi_j$ is smoothed around $\tau_{lj}$, with the exact 
behavior dependent on the form of the energy cut-off. For some
special form of the cut-off, \cite{b15} the shape of the 
instantons for short times can be found from the equations of 
motion. It minimizes the instanton contribution to the action in 
the absence of the long-time interactions and determines the 
instanton tunneling amplitudes. The amplitudes are, in addition,
modified by quantum corrections and we take them as some 
unspecified parameters $W_j/(2 \pi \alpha)$.

The low-energy part of the action determines the 
instanton-instanton interaction and can be found by substitution 
of the long-time asymptotic form of $\Phi_j(\tau)$, i.e., 
$\Phi_j(\tau)= 2 \pi \times \mbox{integer}+\sum_l e_{lj}
2\pi \Theta(\tau-\tau_{lj})$ into the $\Phi_j$ action 
(\ref{e15}). Indeed, the constant part of $\Phi_j(\tau)$ is 
irrelevant, and as we will see later, only the trajectories with 
the same initial and final values of $\Phi_j(\tau)$ for each 
contact $j$, $\sum_l e_{lj}=0$, are important. The Fourier 
transform $\Phi_j(\omega)=\sqrt{T} \int_0^{1/T} d\tau e^{i\omega 
\tau} \Phi_j(\tau)$ of such ``neutral'' trajectories is:
\begin{equation}
\Phi_j(\omega) = \sum_l e_{lj} \frac{2\pi i \sqrt{T}}{\omega}
e^{i\omega \tau_l} \, ,
\label{e20} \end{equation}
and combining this equation with Eq.~(\ref{e15}) we see that
the low-energy part of the action has the form of the pair-wise
instanton interaction with each pair of instantons with indices
$lj$ and $ki$ contributing the term
\begin{equation}
{\cal S}_{int}(\tau_l,\tau_k) = e_{lj}e_{ki} T \sum_{\omega}
\left(\frac{2\pi }{\omega} \right)^2 \hat{K}^{-1}_{ij} (\omega)
e^{i\omega (\tau_l-\tau_k)} .
\label{e21} \end{equation}

It is convenient to formulate the instanton dynamics in terms
of the fields $\Theta_j$, $j=1\, ...\, n$, that are defined as 
dual to $\Phi_j$, i.e., satisfy the commutation relations:
\begin{equation}
[\Theta_{j},\Phi_{j'}]=2\pi i \delta_{jj'} \, .
\label{e22} \end{equation}
In terms of these fields, each instanton tunneling is generated 
by the operator
\begin{equation}
\frac{W_j}{2 \pi \alpha} \mbox{exp}\{i e_{lj} \Theta_j(\tau_l)\}
\label{e23} \end{equation}
which shifts $\Phi_j$ by $2\pi e_{lj}$. The low-energy part of
the action containing instanton-instanton interaction (\ref{e21})
can be understood then as arising from Gaussian fluctuations of
the dual fields:
\[ e^{-{\cal S}_{int}} = e^{- e_{lj}e_{ki} 
\langle\Theta_i (\tau_k) \Theta_j(\tau_l) \rangle/2 }  \, ,  \] 
and Eq.~(\ref{e21}) gives the correlators of these fields:
\[ \langle \Theta_i (-\omega) \Theta_j(\omega)\rangle = (2\pi/
\omega)^2 \hat{K}_{ij}^{-1}(\omega) \, . \]

To find these correlators explicitly, we need to invert the
matrix $\hat{K}$ defined in Eq.~(\ref{e16}). This task is not
difficult, since the matrix has a triangular form: for
negative frequencies, $\hat{K}$ and $\hat{K}^{-1}$ are
upper-triangular, while for positive frequencies they are
lower-triangular. In both situations, one can find
$\hat{K}^{-1}$ by writing equations for the matrix elements
$\hat{K}^{-1}_{ij}$ as recurrence relations in terms of the
``distance'' from the diagonal and solving them starting with
the diagonal elements. For negative frequencies, the result is:
\begin{eqnarray}
\langle \Theta_i (-\omega) \Theta_j(\omega)\rangle =
(2\pi/\omega)^2 \hat{K}^{-1}_{ij} =  \;\;\;\;\; \;\;\;\;\;
\nonumber \\ = {2 \pi \over |\omega|}\left[{2 \over
\lambda_i^2} \delta_{ij} - \Theta( \mbox{sgn(i-j)}\omega)B_{ij}
e^{-|\omega t_{ij}|} (1- \delta_{ij}) \right] \, ,
\label{e25} \end{eqnarray}
\begin{equation}
B_{ij} = {4\over \nu_0 \lambda_i^2\lambda_j^2}
\prod_{i<k<j}\left(1- {2\over \nu_0 \lambda_k^2} \right) \, .
\label{e26} \end{equation}
For positive frequencies, Eq.~(\ref{e25}) remains valid if we 
define the matrix $B_{ij}$ for $i>j$ (below the diagonal) by the
condition $B_{ij}=B_{ji}$. Qualitatively, as will be discussed
in more details below, the dual fields $\Theta_j$ describe
quasiparticles backscattered at the point contacts, and the 
matrix $B_{ij}$ (\ref{e26}) determines the exchange statistics 
of these quasiparticles.

Similarly to the case of electron tunneling model where the
correlators (\ref{e16}) give the kinetic part (\ref{e15}) of
the action, the correlators (\ref{e25}) determine the kinetic 
part of the action of the dual model of quasiparticle tunneling:
\begin{equation}
\bar{{\cal S}}_0 =\frac{1}{2} \Theta (\omega/2\pi)^2 \hat{K} 
\Theta \, .
\label{e27} \end{equation}
One more remark is that the $\Theta_j$-correlators (\ref{e25})
show that the condition of neutrality of instanton tunneling
trajectories, $\sum_l e_{lj}=0$, that was used above, is satisfied
at each point contact $j$. Indeed, the form (\ref{e25}) of the
correlators implies that interaction between instantons diverges
at low energies and only the trajectories that satisfy the
neutrality condition for each $j$ have finite action and
contribute to the evolution of the system.

\subsection{Dual chiral fields and Klein factors}

One can see directly that the expansion of the system
propagator in the number of instanton tunneling events can
be generated as the expansion in the instanton tunneling
Lagrangian composed of the tunneling operators (\ref{e23}).
This means that the dual model of quasiparticle tunneling can
be formulated in complete analogy to the direct model of
electron tunneling. Comparing the correlators (\ref{e25}) with
those from Eq.~(\ref{e16}), one can notice that similarly to the
$\Phi$ fields (\ref{e14}), $\Theta_j$ can be represented as the
sum
\begin{equation}
\Theta_j= \frac{2}{\lambda_j} \theta_j +\bar{\eta}_j
\label{e28} \end{equation}
of the fields with, respectively, chiral and pure statistical
correlators. Equation (\ref{e25}) implies then that the chiral
correlator of the fields $\theta_j$ for $i\neq j$ is:
\begin{equation}
\langle \theta_i\theta_j \rangle =-(\lambda_i \lambda_j/4)
B_{ij} g(x_i-x_j,\omega) \, .
\label{e29} \end{equation}
The representation (\ref{e28}) of $\Theta_j$ allows us to
write the instanton tunneling Lagrangian in the form analogous
to Eq.~(\ref{e10}):
\begin{equation}
\bar{{\cal L}}_{tunn}= \sum_{j=1}^{n} [{W_j\over 2\pi \alpha}
\bar{F}_j \exp \{i \frac{2}{\lambda_j} \theta_j(t) \}+ h.c.]\, .
\label{e30} \end{equation}
where the Klein factors of the dual model are:
\begin{equation}
\bar{F}_j=e^{i\bar{\eta}_j}, \;\; [\bar{\eta}_i,\bar{\eta}_j]=
-i\pi \mbox{sgn}(i-j)(1- \delta_{ij}) B_{ij}\, .
\label{e31} \end{equation}

Before discussing the physical consequences of the dual model of 
quasiparticle backscattering, we present the derivation of this 
model that is less rigorous than the one that uses instanton 
expansion, but more direct. The approach is based 
on independent application of the known strong-tunneling solution 
for one point contact \cite{b17} to individual contacts of our 
multi-point-contact junction and matching the obtained solutions 
at successive contacts along the edge $\nu_0$ common to all of 
them (see Fig.\ 1). Since the bosonic fields at different point 
contacts interact strongly at low frequencies, the possibility to 
simply match the solutions at successive contacts is by no means 
trivial and represents an important assumption. The instanton
calculation described in the previous subsection can be viewed as 
the proof that this assumption is indeed valid provided one 
chooses correctly the statistical phases of electrons in different 
contacts, which after duality transformation determine the 
statistical phases of backscattered quasiparticles.

In more details, consider the $j$th point contact of our
multi-contact model. Strong-tunneling solution (reviewed in 
the Appendix) can be described as application of the Dirichlet 
boundary condition to the tunneling field $\varphi_j(x)$ which 
localizes this field at the minimum of the tunnel potential at 
the point of contact $j$. ``Unfolded'' form of this condition 
\cite{b18} implies free propagation of the two fields constructed 
from the original bosonic modes $\phi_{j}(x)$ and $\phi_{0}(x)$
of the edges forming this contact. One is the field
\begin{equation}
\theta_j(x)=-\mbox{sgn}(x-x_j) \varphi_j (x)
\label{e35} \end{equation}
dual to the tunneling field $\varphi_j(x)$, while the other,
$\tilde{\varphi}_j (x)$ is the combination of $\phi_j(x)$
and $\phi_0(x)$ orthogonal to the tunneling field, i.e.,
\begin{equation}
\varphi_j= \frac{1}{\lambda_j} ({\phi_0 \over
\sqrt{\nu_0}} - {\phi_j \over \sqrt{\nu_j} }), \;\;
\tilde{\varphi}_j= \frac{1}{\lambda_j} ( {\phi_0 \over
\sqrt{\nu_j}} + {\phi_j \over \sqrt{\nu_0} } ).
\label{e32} \end{equation}
(We use the same notation for the dual field $\theta_j$
defined by Eq.~(\ref{e35}) as in Eq.~(\ref{e28}) in anticipation
of the fact, proven later, that these two fields indeed
coincide.) Free evolution of the field $\theta_j(x)$ (\ref{e35})
is equivalent to the statement that the incoming tunneling field
changes sign at the contact $x=x_j$. Combining this change of
sign with Eq.~(\ref{e32}), we see that the transformation of the
incoming fields $\phi_{0}(x)$ and $\phi_{j}(x)$ at $x<x_j$ into
outgoing fields at $x>x_j$ (Fig.~2a) can be described
\cite{b17} with the 2-by-2 matrix $\hat{P}^{(j)}$:
\begin{equation}
(\phi_0^{(out)},\phi_j^{(out)})^{T}=\hat{P}^{(j)} (\phi_0^{(in)},
\phi_j^{(in)})^{T}
\label{e33} \end{equation}
with the matrix elements
\begin{equation}
\hat{P}^{(j)}_{00}= -\hat{P}^{(j)}_{jj}={\nu_0- \nu_j\over \nu_0
+ \nu_j} \, , \;\; \hat{P}^{(j)}_{0j} = \hat{P}^{(j)}_{j0}=
{2\sqrt{\nu_0 \nu_j} \over \nu_0 +\nu_j} \, .
\label{e34} \end{equation}

\begin{figure}[htb]
\setlength{\unitlength}{1.0in}
\begin{picture}(3.,2.)
\put(.05,.1){\epsfxsize=3.1in\epsfbox{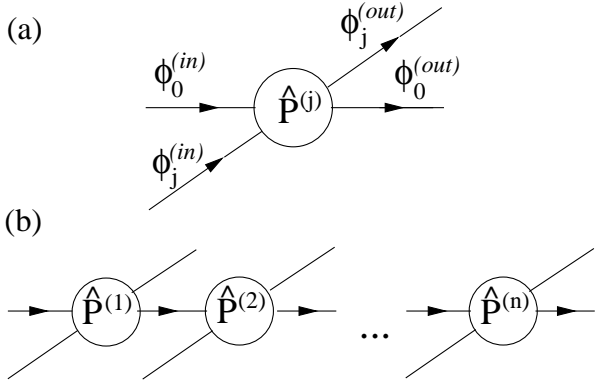}}
\end{picture}
\caption{Diagram of the strong-coupling scattering of the edge
states in the multi-point-contact junction considered in this 
work: (a) transformation of the bosonic edge-state fields 
\protect (\ref{e33}) in one individual point contact $j$; and 
(b) transformation of the fields in the $n$-point junction as a 
whole: the field $\phi_0$ undergoes successive transformations 
in the course of propagation along the edge $\nu_0$.}
\end{figure}

A natural way to combine transformations (\ref{e33}) at the 
neighboring contacts is to {\em assume} that the field 
$\phi_0^{(out)}$ coming out of each contact along the edge 
$\nu_0$ serves as the incoming field $\phi_0^{(in)}$ for the 
next point contact downstream this edge (Fig.~2b). Under this 
assumption (validity of which implies the non-trivial mutual 
statistics of backscattered quasiparticles), description of
the overall transformation of the incoming bosonic modes
in our multi-point-contact junction can be obtained by
extending each of the matrices $\hat{P}^{(j)}$ trivially
to the $n+1$ by $n+1$ matrix: the fields $\phi_i$ with
$i\neq j$ are constant in $\hat{P}^{(j)}$, and
multiplying this matrices. The full transformation of
the incoming fields $\phi^{(in)}=(\phi_0,\phi_1,...\,
,\phi_n)^T$ into the outgoing fields (see Fig.~2b) is
described then as:
\begin{equation}
\phi^{(out)}=\hat{S} \phi^{(in)}, \;\;\;\; \hat{S}=
\hat{P}^{(n)}\hat{P}^{(n-1)}\cdot ...\cdot \hat{P}^{(1)}.
\label{e36} \end{equation}

Such a combination of successive transformations of the fields
at the point contacts together with the free propagation of the
field $\phi_0$ between the contacts implies that the correlator
of the fields $\theta_j(x)$ defined by Eqs.~(\ref{e35}) and
(\ref{e32}) at different contacts is:
\[ \langle \theta_i(x_i)\theta_j(x_j)\rangle =
-\frac{1}{\nu_0 \lambda_i \lambda_j}  g(x_i-x_j, \omega)
\prod_{i<k<j} \hat{P}^{(k)}_{00} \, .  \] 
Equations (\ref{e9}), (\ref{e26}), and (\ref{e34}) show that 
this correlator coincides with the correlator (\ref{e29}).
According to Eq.~(\ref{e35}), the correlator $\langle
\theta_j\theta_j\rangle$ at the same contact is $\langle
\theta_j\theta_j\rangle= \langle\varphi_j\varphi_j \rangle= 
\pi/|\omega|$ and also agrees with Eqs.~(\ref{e28}) and 
(\ref{e25}). This means that the chiral fields $\theta_j$ 
introduced in Eqs.~(\ref{e28}) and (\ref{e30}) as the fields 
describing the quasiparticles backscattered at the point 
contacts coincide with the dual fields $\theta_j(x_j)$ from
Eq.~(\ref{e35}) matched directly between the contacts.
This coincidence is non-trivial and is only correct if
the mutual statistics of quasiparticles backscattered at
different point contacts is accounted for by inclusion of 
the Klein factors $\bar{F}_j$ (\ref{e31}). The statistics, 
however, does not affect the scaling dimensions of the
quasiparticle tunneling operators, which have the same
values $2/\lambda^2_j$ as for one independent point contact 
\cite{b16}. This means that the dual model of strong 
tunneling in the multi-point-contact junction constructed in 
this work is stable for large temperatures and/or voltages 
applied to all contacts, and the strong-tunneling limit 
is always reached at sufficiently large energies.

\subsection{Interpretation of the quasiparticle statistics
and calculation of the outgoing currents and shot noise}

In this subsection, we use the approach developed above
to calculate the transport properties of the
multi-point-contact junction in the limit of strong
tunneling. Application of the voltage $V_l$ to the $l$th
edge can be described simply as a shift of the incoming
field (see, e.g., the Appendix)
\begin{equation}
\phi^{(in)}_l \to \phi^{(in)}_l - \sqrt{\nu_l} V_l t\, ,
\label{e39} \end{equation} which can be understood as
the evolution of the quantum-mechanical phase of the
electron operator $\psi_l$ (\ref{e2}) of the $l$th edge.
(In this work, we discuss only the dc transport
properties of the junction biased by constant voltages
$V_l$.) Numerical shift of the fields $\phi_l$ does not affect 
their properties as quantum operators, and all the results for 
$\phi_l$ discussed above remain valid in the presence of 
finite bias voltages.

We consider first the situation with {\em no backscattering}. 
Since the current $j_l$ in the $l$th edge is related to 
$\phi_l$ as $j_l(x,t) =-\sqrt{\nu_l} \partial_t \phi_l(x,t)/2 
\pi$, the outgoing currents are linear in the
applied voltages and one can define the matrix of
conductances through the current induced in the $l$th edge by 
the voltage applied to the $k$th edge: $\hat{G}_{lk}= j_l/V_k$. 
The transformation of the fields $\phi_l$ according to 
Eq.~(\ref{e36}) means that the conductance matrix is:
\[ \hat{G}_{lk}=\sigma_0\sqrt{\nu_l\nu_k} \hat{S}_{lk}\, ,\]
where $\sigma_0$ is the universal free-electron conductance
$e^2/h$ equal to $1/2\pi$ in our units $e=\hbar=1$. Direct
substitution of the matrices (\ref{e34}) into this expression
gives
\begin{eqnarray}
G_{00}&=& \nu_0 \prod_1^n {\nu_0-\nu_l \over \nu_0+\nu_l} \ ,
\nonumber  \\
G_{jj}&=& \nu_j {\nu_j-\nu_0 \over \nu_0+\nu_j} \ , \;\;\;\;
G_{ij}=0 \;\;  \mbox{for} \ \  j>i\geq1, \label{e40} \\
G_{ij}&=&{4 \nu_0 \nu_i \nu_j \over (\nu+\nu_i)(\nu_0+\nu_j)}
\prod_{j<l<i}{\nu_0-\nu_l \over \nu_0+\nu_l} \;\; \mbox{for} \
1\le j <i \le n, \nonumber \\
G_{0j}&=& {2 \nu_0 \nu_j \over \nu_0+\nu_j} \prod_{l>j}^n
{\nu_0-\nu_l \over \nu_0+\nu_l} \ , \;\;\;\;
G_{j0}={2 \nu_0 \nu_j \over \nu_0+\nu_j}
\prod_{l<j} {\nu_0-\nu_l \over \nu_0+\nu_l} \, , \nonumber
\end{eqnarray}
where $G_{ij}$ denotes the conductance $\hat{G}_{ij}$ in units
of $\sigma_0$. Charge conservation requires that $\sum_l 
G_{lj}= \nu_j$, the condition that is indeed satisfied by 
Eqs.~(\ref{e40}).

Equations (\ref{e40}) describe a flow of charge along the edges 
in which the difference of currents in the edges $0$ and $j$ is 
redistributed at the $j$th point contact according to its 
tunneling conductance $\sigma_0 (\nu_j-G_{jj})=2 \sigma_0 \nu_0 
\nu_j/(\nu_0+\nu_j)$. Such a description of current flow in the 
system of several point contacts was first obtained in
Ref.~\onlinecite{b12} under the assumption that quantum 
coherence of electron propagation along the common edge is 
suppressed, and electron distribution is equilibrated between 
the successive point contacts. The calculations presented above 
show that this description of the current flow remains valid in 
the quantum-coherent regime, when any external decoherence 
mechanisms are absent.

We now turn to the regime of {\em finite backscattering}. As 
discussed above, each elementary process of backscattering can 
be described as instanton tunneling generated by the operator 
$\propto  \exp \{ \pm i 2 \theta_j /\lambda_j\}$ [see 
Eq.~(\ref{e30})]. If the tunneling occurs in the point contact 
$j$, the associated shift of the field $\phi_j$ corresponds to
the transfer of charge $q_j=2/\lambda_j^2= 2\nu_0\nu_j/(\nu_0+ 
\nu_j)$ (see, e.g., the Appendix). In the multi-point-contact 
junction, quasiparticle of charge $q_j$ produced in the $0$th 
edge at the point $x_j$ by instanton tunneling propagates 
along this edge and is split between edges $i$ with $i>j$ and 
the edge $0$ according to the conductance matrix. Distribution 
of charges in backscattering processes that results from such 
a ``current branching'' at the point contacts enables us to 
interpret the quasiparticle Klein factors (\ref{e31}) of the 
dual model as exchange statistics of charges generated in the 
edge $0$. Indeed, the absence of the energy gap at the edge of 
a FQHL makes it possible to produce arbitrary edge 
excitations. For an isolated edge $0$, the exchange statistics 
of two excitations with charges $p_1$ and $p_2$ is set by its 
filling factor $\nu_0$ to be $p_1p_2/\nu_0$. The two 
quasiparticles which tunnel in the nearest-neighbor contacts 
create charges $q_j$ and $q_{j+1}$ in the edge $0$ and the
statistical angle for these quasiparticles, $2\pi
q_jq_{j+1}/\nu_0$ is easily seen to indeed coincide with
the statistical angle $2\pi B_{j,j+1}$ determined by the
Klein factors (\ref{e31}) through the statistical matrix
$B_{ij}$ (\ref{e26}). If the quasiparticles tunnel at
the point contacts that are not nearest neighbors, one
needs to take into account that the charge created in
the edge $0$ in the vicinity of the contact $i$ by
quasiparticle tunneling at the contact $j<i$ is smaller
than $q_j$ by the factor
\[ r\equiv \prod_{j<l<i}[(\nu_0-\nu_l)/(\nu_0+\nu_l)] \]
due to charge splitting at the intermediate point contacts.
Equation (\ref{e26}) shows then that the matrix $B_{ij}$ can 
be written as $B_{ij}=rq_iq_j/\nu_0$, i.e., the statistics 
of quasiparticles can still be interpreted as the exchange
statistics of charges in the common edge $0$.

This interpretation remains valid if some of the filling
factors $\nu_l$ are equal to each other. For instance, if the
edges $j$ and $j+1$ have the same filling factor $\nu_0$ as 
the edge $0$, the charge and statistics of the backscattered
quasiparticles coincide with those of Laughlin's 
quasiparticles: $q_j=B_{j,j+1}=\nu_0$, in agreement with 
previous results \cite{b19}. Our discussion here shows that 
in general the strong coupling between non-identical edges 
should produce quasiparticles with both charge and statistics 
that are different from those of Laughlin's quasiparticles in 
the bulk.

To find the backscattered currents quantitatively, we
calculate first the charges injected into each edge by
quasiparticle tunneling at the $j$th contact. Combining the two 
facts, that the fraction of the incoming current in the $j$th 
edge that goes out into the $i$th edge is equal to 
$G_{ij}/\nu_j$, and that the quasiparticle charge $q_j$ is 
created in the process of backscattering directly in the $0$th 
edge, we see from Eqs.~(\ref{e40}) that each backscattering 
event in the contact $j$ creates the charge equal to $G_{ij}$ 
in the $i$th edge, $i\neq j$. This means that the 
backscattering current $I^{(bsc)}_l$ in the $l$th outgoing edge 
($l=0\, ...\,n$) is related to the rate $I^{(0)}_j$ of 
quasiparticle backscattering in the $j$th contact as:
\begin{equation}
I^{(bsc)}_l = \sum_{j=1}^n \Delta G_{lj} I^{(0)}_j \, , \;\;\;
\Delta G_{lj}\equiv G_{lj}- \nu_j \delta_{lj} \, .
\label{e41} \end{equation}

To find the rates $I^{(0)}_j$, we need to calculate the
distribution of bias voltages for quasiparticles. In
contrast to finding the bias (\ref{e39}) for electron
tunneling, this task is not completely trivial, since in
the strong tunneling limit, the bias voltage $V_l$
applied to the $l$th edge affects the quasiparticle bias
$\bar{V}_j$ in all point contacts. The bias $\bar{V}_j$
can be obtained as the voltage-induced shift of the dual
fields $\theta_j$ in the arguments of exponents in the
tunneling Lagrangian (\ref{e30}). We find this shift
from the linear relation between the incoming fields
$\phi^{(in)}_l$ and the fields $\theta_j$ (\ref{e35}).
Extending the set of $n$ fields $\theta_j$ to the
$(n+1)$-component vector $\theta \equiv (\phi^{(out)}_0,
\theta_1, \theta_2, ...,\theta_n)^T$ we can write this
linear relation similarly to Eq.~(\ref{e36}):
\begin{equation}
\theta =\hat{S}' \phi^{(in)}\, , \;\;\;\; \hat{S}'=
\hat{T}^{(n)}\hat{T}^{(n-1)}\cdot ...\cdot\hat{T}^{(1)},
\label{e42} \end{equation}
where the matrices $\hat{T}^{(j)}$ describe transformations 
of the incoming fields into $(\phi^{(out)}_0, \theta_j)$ in
the individual point contacts. The part of $\hat{T}^{(j)}$
related to $\phi^{(out)}_0$ is the same as in $\hat{P}^{(j)}$
(\ref{e33}), while the part related to $\theta_j$ follows
directly from Eqs.~(\ref{e35}) and (\ref{e32}):
\begin{eqnarray}
\hat{T}^{(j)}_{j0}=\left({\nu_j \over \nu_0 +\nu_j}
\right)^{1/2}, \;\; \hat{T}^{(j)}_{jj}=-\left(
{\nu_0 \over \nu_0 +\nu_j}\right)^{1/2} , \nonumber \\
\hat{T}^{(j)}_{00}=\hat{P}^{(j)}_{00}\, , \;\;\;
\hat{T}^{(j)}_{0j}=\hat{P}^{(j)}_{0j} \, .
\;\;\;\;\;\;\;\;\;\;\;\;\;\;\;\;\;\;
\label{e43} \end{eqnarray}

Since the voltages $V_l$ shift the incoming fields according
to Eq.~(\ref{e39}), the transformation (\ref{e42}) shows that
the voltage-induced shift of the quasiparticle fields $\theta_j$
is
\[ \theta_j \to \theta_j - \sum_l \hat{S}'_{jl}
\sqrt{\nu_l} V_l t \, .  \]
Obtaining the matrix elements $\hat{S}'_{jl}$ explicitly from
Eqs.~(\ref{e43}) and (\ref{e34}) we see that the voltage bias
in the quasiparticle tunneling operator in Eq.~(\ref{e30}) is:
\begin{equation}
{2 \theta_j \over \lambda_j}  \to {2 \theta_j \over \lambda_j}
 -\bar{V}_jt\, , \;\;\; \bar{V}_j= \sum_l \Delta
G_{jl} V_l \, . \label{e44} \end{equation} Inserting
this result into the expression for the rate of
quasiparticle tunneling  from the $0$th edge at the
$j$th contact that follows from Eq.~(\ref{e30}) we get:
\begin{equation}
I^{(0)}_j = - {W_j \over \pi \alpha} \sin \left({2 \over
\lambda_j} \theta_j +\bar{\eta}_j- \bar{V}_jt \right) \,
. \label{e45} \end{equation}

Finally, combining the strong-coupling branching of the currents
according to the conductance matrix (\ref{e40}) with the
backscattering contribution to the current, we see that the
average {\em outgoing current} $\langle I_l\rangle$ in the $l$th
edge is:
\begin{equation}
\langle I_l\rangle = \sigma_0 \sum_k G_{kl} V_l + \langle
I^{(bsc)}_l \rangle\, .
\label{e46} \end{equation}
In the limit of strong tunneling, the backscattering is weak 
and the average quasiparticle tunneling rate $\langle I^{(0)}_j 
\rangle$ can be evaluated in the lowest order of perturbation 
theory in backscattering (\ref{e30}). For large quasiparticle 
bias voltages $\bar{V}_j$ one gets: 
\[\langle I^{(0)}_j \rangle = {W_j^2 D \over 2 \pi v^2 \Gamma 
(4/\lambda_j^2 )} \mbox{sgn}(\bar{V_j})|\bar{V}_j/D|^{ 
(2/\lambda_j)^2-1}  \, . \] 
This expression combined with Eqs.~(\ref{e41}) and (\ref{e46}) 
describes how the average outgoing current in the $l$th edge
approaches its linear large-voltage asymptotics.

In this regime of weak backscattering, the quasiparticle
tunneling produces regular {\em shot noise}, which at low
temperatures is the only source of noise of the outgoing
currents.\cite{b20} This mean that the correlators of the
outgoing currents $\langle I_l\rangle$ at zero frequency
can be expressed as:
\begin{equation}
\langle \{I_l,I_k\} \rangle|_{\omega \to
0}=2\sum_{j=1}^n \Delta G_{lj} \Delta G_{kj} |\langle
I^{(0)}_j \rangle| \, , \label{e47}
\end{equation}
where $\{... , ...\}$ denotes the
anticommutator. In general, the current noise
(\ref{e47}) is not proportional to the average
backscattered current, unless the contribution to
backscattering of one point contact is dominant. If the
contact $m$ does dominate, the shot noise of the
outgoing current in the $l$th edge is characterized by
the fractional charge equal to $\Delta G_{lm}$,
\begin{equation}
q^* \equiv \left|\frac{\langle \{I_l,I_l\}
\rangle|_{\omega \to 0}}{ 2\langle I^{(bsc)}_l \rangle}
\right| = |\Delta G_{lm}| \, , \label{e48}
\end{equation} i.e., the charge is proportional to the
fraction of the quasiparticle charge which reaches the
$l$th edge in the current branching process. The shot
noise with the charge (\ref{e48}), generated at one
point contact and fractionally split into another
contact, is direct manifestation of coherent
propagation of electrons between the point contacts.
Equilibration of electron distribution between the
contacts would leave in each edge only the shot noise
generated by the backscattering at the point contact
formed by this edge.

\section{Tunneling between multi-mode Fermi-liquid 
reservoir and FQHL}

In this Section, we consider the situation with one specific 
choice of the filling factors, $\nu_j=1$, for $j=1 \div n$, 
in the general model of the multi-point-contact junction 
(Fig.~1). The multi-point-contact junction in this regime has 
been used before \cite{b11,b12,b13} to model an interface
between the multi-mode Fermi-liquid reservoir and the 
FQHL. In the limit of strong tunneling, the
reservoir-FQHL interface has all the features of a
regular Ohmic contact: linear current-voltage characteristics 
and no shot noise, and establishes equilibrium between the 
reservoir and the edge of the FQHL when the number $n$ of 
reservoir modes is large. Results of this section describe 
corrections to the Ohmic behavior caused by the weak 
quasiparticle backscattering which manifests itself through 
the shot noise and non-linear corrections to current.

In the case of reservoir-FQHL interface, the $\nu=1$ edges of
the general model of Fig.~1 represent one-dimensional
free-electron scattering modes of the $n$-mode Fermi-liquid 
reservoir. When a voltage $V$ is applied to the FQHL edge, the 
tunneling current $I_T$ that flows between the reservoir and 
the edge can be found as the sum of the outgoing currents [see
Eqs.~(\ref{e41})and (\ref{e46})] in the free-electron modes
$j=1 \div n$:
\begin{equation}
I_T=-\sigma_0 \Delta G_{00} V - \sum_{j=1}^n G_{0j}
I^{(0)}_j (G_{j0}V) \, . 
\label{e50} \end{equation} 
The argument of the $j$th rate of backscattering $I^{(0)}_j$
here represents the quasiparticle bias voltage (\ref{e44}) 
in the $j$th contact.

Perturbation theory in quasiparticle tunneling
(\ref{e30}) of the dual model is justified when either
the temperature or all quasiparticle bias voltages are
larger than the energy scale $T_X \simeq \mbox{max}_j
(W_j/v)^{1/(1-2/\lambda^2)} D$ of the crossover to
strong tunneling in all point contacts. The condition on
the bias voltages is always satisfied if the applied
voltage $V$ is sufficiently large: $|V|
[(1-\nu_0)/(1+\nu_0)]^n
>T_X$.

Tunneling at large temperatures can be characterized by the
{\em linear conductance} $G$ of the reservoir-edge interface.
We calculate this conductance by first obtaining the average
rates $\langle I^{(0)}_j \rangle$ of backscattering in each
contact in the first non-vanishing order of perturbation theory
in tunneling (\ref{e30}) and then summing them according to
Eq.~({\ref{e50}). We find that with increasing temperature, 
conductance $G$ approaches its saturation value $G_n$,
\begin{equation}
G_n=\sigma_0 \nu_0 [1-(q-1)^n]\, , \;\;\; q=\frac{2
\nu_0}{1+\nu_0} =\frac{2}{\lambda^2}  \, ,
\label{e51} \end{equation}
reached in the absence of backscattering, as
\begin{equation}
G=G_n-\frac{q^2  (q-1)^{n-1}}{4 \sqrt{
\pi}}\frac{\Gamma(q)}{ \Gamma(q+1/2)} \left(\frac{\pi
T}{D}\right)^{2(q-1)}\sum_j{W_j^2\over v^2}\, .
\label{e52} \end{equation} 
Since $q<1$, the first-order perturbation correction to 
$G_n$ (\ref{e52}) vanishes with increasing temperature as 
a power of temperature, $G-G_n \propto 1/T^{2(1-q)}$. In 
the limit of large number of point contacts, $n \to \infty$, 
this power-law correction is, however, suppressed by the 
same small factor $(1-q)^n$ which characterizes how
$G_n$ approaches its equilibration value $\sigma_0 \nu$
at $n\to \infty$. One can show that correction to $G_n$
in the next non-vanishing order of the perturbation
theory in backscattering is also suppressed by this
factor. This suggests that $G$ approaches the saturation
value faster than any negative power of $T$.

Following the same step as for the linear conductance
(\ref{e52}) we obtain the average {\em tunneling current} in
the large-voltage regime:
\[ \langle I_T \rangle =G_n V+ \frac{(-1)^n q^{2q}V}{2 \pi
\Gamma(2q)} (V/D)^{2(q-1)} (1-q)^{(n-1)(2q-1)} \]

\vspace{-2ex}

\begin{equation}
\times  \sum_j  (W_j/v)^2 (1-q)^{2(n-j)(1-q)} \, .
\label{e53} \end{equation} For large $n$, the main
contribution to the sum over $j$ in this equation comes
from the contacts at the end of the junction, where $j
\simeq n$. The overall magnitude of the average
backscattered current (\ref{e53}) in the perturbative
regime of weak backscattering is proportional to
$(1-q)^{(n-1)(2q-1)}$, and for $\nu_0 \ge 1/3$, does not
vanish when $n \to \infty$. If $\nu_0=1/3$, the exponent
in this dependence is zero:  $2q-1=0$, while for smaller
$\nu_0$ the exponent is negative and makes the absolute
value of the backscattered part of current (\ref{e53}) a
growing function $n$. Equations~(\ref{e52}) and(\ref{e53}) 
show also that the sign of the backscattering
contribution to the tunnel current oscillates with the
parity of the number $n$ of point contacts, with
backscattering increasing the tunnel current for even $n$.

The {\em shot noise} of the tunneling current is given
by the zero-frequency correlator $\langle \{I_T,I_T\}
\rangle$ and similarly to Eq.~(\ref{e47}) is generated
by the backscattering part of the current (\ref{e50}):
\begin{equation}
\langle \{I_T,I_T\} \rangle= \sum_{i,j=1}^n G_{0i}
G_{0j}\langle \{I^{(0)}_i,I^{(0)}_j\} \rangle \, .
\label{e54} \end{equation} 
In the lowest non-vanishing
order in the backscattering, the correlators on the
right-hand-side of Eq.~(\ref{e54}) are not equal to zero
only for $i=j$. Non-zero diagonal correlators are given
by the average backscattered currents, as in
Eq.~(\ref{e47}). In this case, summing the contributions
$2 \langle I^{(0)}_j \rangle$ to noise from each of the
point contacts according to Eq.~(\ref{e54}) we get
\[ \langle \{I_T,I_T\} \rangle = \frac{q^{2q+1}V}{\pi 
\Gamma(2q)} (V/D)^{2(q-1)} (1-q)^{(n-1)(2q-1)} \]

\vspace{-2ex}

\begin{equation}
\times  \sum_j(W_j/v)^2 (1-q)^{(n-j)(3-2q)} \, .
\label{e55} \end{equation} 
Similar to the average backscattered current, the main 
contribution to the shot noise comes from few contacts with 
$j \sim n$. If the tunneling amplitudes of these contacts 
are approximately equal so that we can neglect their 
variations, $W^2_n \simeq W^2_{n-1} \simeq ...$, the ratio 
between the noise and the average backscattered current is
\begin{equation}
\left| \frac{\langle \{I_T,I_T\} \rangle }{2(\langle I_T
\rangle +\sigma_0 \Delta G_{00} V)} \right| =
q\frac{1-(1-q)^{2(1-q)} }{1-(1-q)^{3-2q}} \, .
\label{e56} \end{equation} We see that in a uniform
junction with large number $n$ of modes, the splitting
of excitations at the point contacts complicates the
relation between the backscattered charge and the shot
noise.  For instance, if $\nu_0=1/3$, by
a numerical coincidence Eq.~(\ref{e56}) gives the
noise-to-current ratio also equal to $1/3$, while the
backscattered charge is $q=1/2$. For smaller $\nu_0$,
the ratio (\ref{e56}) does not correspond directly to
either $q$ or $\nu_0$.

\section{Conclusion}

To summarize, this work provides a description of quasiparticle 
tunneling in the junction with geometry shown in Fig.~1, in 
which several point contacts are formed between single-mode FQHL 
edges with in general different filling factors $\nu_j=1/odd$ and 
one common single-mode edge with filling factor $\nu_0$. This
description extends to finite reflection in the contacts our 
previous theory \cite{b13} of strong electron tunneling in such 
a junction. New quasiparticle physics emerging form the 
multi-point-contact geometry is coherent splitting of 
quasiparticles at all point contacts downstream the common edge 
$\nu_0$ from the point contact where the quasiparticles have 
been created. Quasiparticles are split at each contact in 
precisely the same way as the incoming edge currents in the 
strong tunneling limit. This means that the branching ratios 
of the edge currents determine the quasiparticle charge and 
statistics. The statistics is given by the permutation 
relations between the backscattering operators producing 
quasiparticles and makes the whole process of coherent 
splitting at successive point contacts causal. The 
quasiparticle charges determine the magnitude of the shot 
noise of the outgoing current in each edge. In the non-uniform 
junctions, in which one point contact dominates the 
backscattering, the shot noise in any given edge is directly 
related to the charge split in this edge in the branching process 
- see Eq.~(\ref{e48}). In uniform junctions, the charge-noise
relation is less direct [see Eq.~(\ref{e56}) and the example 
discussed next to it] because of averaging over different 
point contacts. 

In the case of a junction between the Fermi-liquid reservoir 
and the FQHL edge, $\nu_j=1$ for $j=1 \div n$, the quasiparticle 
backscattering gives corrections to the reservoir-edge equilibration 
reached with increasing number of reservoir modes, $n\to \infty$, in 
the absence of backscattering. For temperatures $T$ much larger than 
the bias voltage $V$, these corrections are uniformly suppressed by 
the factor $[ (\nu-1)/(\nu+1)]^n$, exponential in the number $n$ of 
modes. In the opposite regime, $T\ll V$, the corrections, i.e. 
the non-linear junction conductance and current shot noise, 
do not vanish, however, even for $n \to \infty$.

Description of the charge and statistics of the edge excitations 
in the multi-point-contact junction, which is developed in this 
work for simple single-mode edges in the junction geometry with 
one common edge for all point contacts, should also be important 
for more complex multi-mode edges and more general junction 
geometries.

\section{Acknowledgments}
The authors would like to acknowledge useful discussion with 
V.J. Goldman and J.K. Jain. This work was supported by the NSA 
and ARDA under ARO contract \# DAAD19-03-1-0126. 

\appendix
\section{One point contact}

In this Appendix, we review the known results \cite{b17,b18} for
strong tunneling in one point contact between two edges with
different filling factors using the approach similar to that 
employed in the main text. For consistency of notations, we number 
the two edges $0$ and $j$, with the filling factors $\nu_0$ and 
$\nu_j$. The Hamiltonian of one point contact can be written as in 
the model described in Secs.~II and III
\begin{eqnarray}
{\cal H}= \int dx \left( \frac{v}{4 \pi} [(\partial_x \phi_0)^2
+(\partial_x \phi_j )^2 ]  + [ V_0 \rho_0 + V_j \rho_j] \right)
\nonumber \\  - u_j
\left. \cos \left( {\phi_0 \over \sqrt{\nu_0}} - {\phi_j \over
\sqrt{\nu_j} }\right)  \right|_{x=0}\, .
\label{a1} \end{eqnarray}
The two differences with Secs.~II and III are that now we 
explicitly include the bias voltages $V_0$ and $V_j$ applied to 
the edges, and omit the Klein factors in the tunnel part of the 
Hamiltonian, since they are irrelevant in the case of one point 
contact. (The Klein factors $\xi_0 \xi_j$ and $\xi_j \xi_0$ in the 
forward and backward tunneling terms commute with each other and 
are constants of motion. This means that they can at most produce 
an irrelevant constant shift of the fields $\phi$.)

The tunneling field $\varphi_j$ is introduced according to
Eqs.~(\ref{e32})  which can be viewed as the rotation in the
space of the fields $\phi$:
\[ \hat{O}= \left( \begin{array}{cc} \displaystyle
\cos\, \vartheta_j \, , & -\sin\, \vartheta_j \\ \sin\,
\vartheta_j\, , & \cos\, \vartheta_j \end{array} \right) \]
by the angle $\vartheta_j$ defined through the relations:
\[ \cos\, \vartheta_j=\sqrt{\nu_j \over \nu_0 +\nu_j}\ , \;\;\;
\sin\, \vartheta_j=\sqrt{\nu_0 \over \nu_0 +\nu_j}\, . \]
Performing this rotation in the Hamiltonian (\ref{a1}), in
particular rotating the ``vector'' of bias voltages
$\sqrt{\nu_0} V_0, \sqrt{\nu_j}V_j$, we see that the
Hamiltonian ${\cal H}_T$ of the tunneling mode $\varphi_j$
can be separated from other terms:
\[ {\cal H}_T= \int \frac{dx}{2 \pi} [\frac{v}{2} (\partial_x
\varphi_j )^2  +\frac{V}{\lambda_j} \partial_x \varphi_j ]
- u_j \cos[ \lambda_j \varphi_j(x=0) ]\, , \]
\begin{equation}
V\equiv V_0 - V_j\, . 
\label{a2} \end{equation}
From now on we omit the index $j$ of all variables. Heisenberg
equation of motion for $\varphi$ obtained from the
Hamiltonian (\ref{a2}) and the commutation relations
(\ref{e7}) is:
\begin{equation}
(\partial_t+ v\partial_x)\varphi(x,t) = - \frac{V(t)}{\lambda}
+ u \pi \lambda \sin [\lambda \varphi(0,t)] \mbox{sgn} (x)\, .
\label{a3} \end{equation}

Separating the bias voltage:
\begin{equation}
\varphi(x,t) = \varphi_0(x,t) - \frac{1}{\lambda}
\int^t dt' V (t')\, ,
\label{a8} \end{equation}
we can write the solution of Eq.~(\ref{a3}) as
\begin{equation}
\varphi_0(x,t)= \frac{2\pi m}{\lambda} +\left\{
\begin{array}{cc} \displaystyle \theta (t-x/v)+\pi
\lambda Q(t), & x<0 \, , \\ \chi (t-x/v)-\pi \lambda
Q(t), & x>0 \, , \end{array} \right.
\label{a4} \end{equation}
where an arbitrary constant $2\pi m/\lambda$ is written
in this form for later convenience, and $Q(t)$ is the
operator of charge transferred through the point contact:
\begin{equation}
\dot{Q}(t)= - u\sin[\lambda \varphi(0,t)] \, .
\label{a5} \end{equation}
The function $\theta (t)$ in Eq.~(\ref{a4}) is an
arbitrary operator which has the meaning of fluctuations
of the field $\varphi(x,t)$ incident on the point contact,
and $\chi$ will be determined from the final solution of
Eq.~(\ref{a3}).

Equation (\ref{a3}) implies that $\varphi(x,t)$ is
continuous as a function of $x$ at $x=0$, i.e.,
\begin{equation}
\varphi_0(0,t)- \frac{2\pi m}{\lambda} =\theta (t)+\pi
\lambda Q(t)=\chi(t)- \pi \lambda Q(t) \, .
\label{a6} \end{equation}
In the strong tunneling limit $u \to \infty$,
we assume that the value of $\varphi(0,t)$ is very close
to one of the minima $2\pi m/\lambda$, $m=0,\pm1, ...$ of
the tunneling term in the Hamiltonian (\ref{a2}) and we can
linearize the sine in Eq.~(\ref{a5}). As we will see in
the end of the calculation, this assumption is indeed correct.
Applying it to the first of the Eqs.~(\ref{a6}), we get
the following simple equation for $\varphi_0(0,t)$:
\begin{equation}
\dot{\varphi}_0 (0,t) = -u \pi \lambda [ \lambda
\varphi_0 (0,t) - \int^{t} dt' V (t') ] +\dot{\theta} \, .
\label{a7} \end{equation}
Solution of this equation under the condition $\dot{\theta},
V \ll u$ appropriate for the strong-tunneling limit gives
\begin{equation}
\varphi(0,t)=\frac{2\pi m}{\lambda} + \frac{1}{u \pi
\lambda^2} [\dot{\theta}(t) - \frac{V(t)}{\lambda}]  \, .
\label{a9} \end{equation}

We see that up to small terms of the order of $1/u$,
$\varphi(0,t)$ is indeed pinned to one of the minima
$2\pi m/\lambda$ of the tunneling energy. Inserting
Eq.~(\ref{a9}) into Eq.~(\ref{a5}) we see that the tunnel
current in the point contact is:
\begin{equation}
\dot{Q}(t) = \sigma_0 \frac{2}{\lambda^2} V(t) -
\frac{\dot{\theta}(t)}{\pi \lambda} \, .
\label{a10} \end{equation}
The first term on the right-hand-side of this equation gives
the strong-tunneling conductance of the point contact, while
the second term represents the tunnel current induced by the
incident field fluctuations. Finally, combining
Eq.~(\ref{a9}) and (\ref{a10}) with the continuity condition
(\ref{a6}) we can determine the transmitted field $\chi(t)$
in Eq.~(\ref{a4}):
\begin{equation}
\chi (t)= \frac{2}{\lambda} \int^{t} dt' V (t') - \theta(t) \, .
\label{a11} \end{equation}
(Equations (\ref{a10}) and (\ref{a11}) include only the terms
that do not vanish in the limit $u\to \infty$.)
Comparison of Eq.~(\ref{a11}) with Eqs.~(\ref{a8}) and
(\ref{a4}) shows explicitly that both the voltage
contribution to $\varphi(x,t)$ and incident fluctuations
$\theta(t)$ change sign at the point contact in the
strong-tunneling limit. This means that the field
$\theta_j(x)$ defined by Eq.~(\ref{e35}) of the main text
indeed propagates freely in the strong tunneling limit.

The last remark is that Eq.~(\ref{a4}) enables us to
obtain directly the charge of backscattered quasiparticles.
Indeed, instanton tunneling that describes the
backscattering changes the number $m$ of the minimum the
field $\varphi(x,t)$ is pinned to, affecting in the
process the charge transfer through the point contact.
Equation (\ref{a4}) shows that the transition from one
tunneling minimum to another changing $m$ by 1 happens
if the transferred charge changes by the amount
\begin{equation}
\delta Q = \frac{2}{\lambda^2} \label{a12}
\end{equation} which coincides with the junction tunnel
conductance in units of $\sigma_0$. One can also see
from Eqs.~(\ref{a4}) and (\ref{a6}) that the large-scale
distribution of the field $\varphi(x,t)$ created by such a 
process of instanton tunneling occurring at time $t_i$ 
coincides essentially, as it should, with the retarded 
Green's function discussed in Sec.~II:
\begin{equation}
\varphi(x,t)= - \frac{2\pi}{\lambda}\Theta(t-t_i)
\mbox{sgn}[x- v(t-t_i)] \, .
\label{a13}
\end{equation}


\begin{thebibliography}{99}

\bibitem{b1} B.I. Halperin, Phys.\ Rev.\ B {\bf 25}, 2185
(1982); P. Streda, J. Kucera, and A.H. MacDonald, Phys.\ 
Rev.\ Lett. {\bf 59}, 1973 (1987); M. Buttiker, Phys.\ Rev.\ B 
{\bf 38}, 9375 (1988).
\bibitem{b4} X.G. Wen, Phys.\ Rev.\ Lett. {\bf 64}, 2206 (1990);
J. Fr\"{o}lich and T. Kerler, Nucl.\ Phys.\ B[FS] {\bf 354}, 369 
(1991); M. Stone and M.P.A. Fisher, Int.\ J.\ Mod.\ Phys.\ B 
{\bf 8}, 2539 (1994); N. Nagaosa and M. Kohmoto, Phys.\ Rev.\ 
Lett. {\bf 75}, 4294 (1995). 
\bibitem{b19} C.L. Kane and M.P.A. Fisher, Phys.\ Rev.\ Lett. 
{\bf 72}, 724 (1994).
\bibitem{b8} V.J. Goldman and B. Su, Science {\bf 267}, 1010 (1995); 
V.J. Goldman, I. Karakurt, J. Liu, and A. Zaslavsky Phys.\ Rev.\ B 
{\bf 64}, 085319 (2001). 
\bibitem{b9} L. Saminadayar, D.C. Glattli, Y. Jin, and B. Etienne, 
Phys.\ Rev.\ Lett. {\bf 79}, 2526 (1997);  R. de-Picciotto, 
M. Reznikov, M. Heiblum, V. Umansky, G. Bunin, D. Mahalu, Nature 
{\bf 389}, 162 (1997); M. Reznikov, R. de Picciotto, T.G. Griffiths, 
M. Heiblum, and V. Umansky, Nature {\bf 399}, 238 (1999).
\bibitem{b21} I. Safi, P. Devillard, and T. Martin, Phys.\ Rev.\ 
Lett. {\bf 86}, 4628 (2001); C.L. Kane, Phys.\ Rev.\ Lett. {\bf 90}, 
226802 (2003); S. Vishveshwara, Phys.\ Rev.\ Lett. {\bf 91}, 
196803 (2003).     
\bibitem{b24} D.V. Averin and V.J. Goldman, Solid State Commun. 
{\bf 121}, 25 (2002). 
\bibitem{b17} N.P. Sandler, C. de C. Chamon, and E. Fradkin,
Phys.\ Rev.\ B {\bf 57}, 12324 (1998); Phys.\ Rev.\ B {\bf 59},
12521 (1999).
\bibitem{b11} C.L. Kane and M.P.A. Fisher, Phys.\ Rev.\ B
{\bf 52}, 17393 (1995).
\bibitem{b12} C. de C. Chamon and E. Fradkin, Phys.\ Rev.\ B
{\bf 56}, 2012 (1997).
\bibitem{b26} A.M. Chang, L.N. Pfeiffer, and K.W. West
{\bf 77}, 2538 (1996); M. Grayson, D.C. Tsui, L.N. Pfeiffer,
K.W. West, and A.M. Chang, Phys.\ Rev.\ Lett. {\bf 80}, 1062 (1998);
A.M. Chang, M.K. Wu, C.C. Chi, L.N. Pfeiffer, and K.W. West,
Phys.\ Rev.\ Lett. {\bf 86}, 143 (2001);
M. Hilke, D.C. Tsui, M. Grayson, L.N. Pfeiffer, and K.W. West,
Phys.\ Rev.\ Lett. {\bf 87}, 186806 (2001);
A.M.Chang and J.E. Cunningham, Surf.\ Science {\bf 229}, 216 (1990).
\bibitem{b13} V.V. Ponomarenko and D.V. Averin, JETP Lett.
{\bf 74}, 87 (2001); Phys.\ Rev.\ B {\bf 67}, 35314 (2003). 
\bibitem{b28} E. Comforti, Y. C. Chung, M. Heiblum, and 
V. Umansky, Phys.\ Rev.\ Lett. {\bf 89}, 066803 (2002); 
E. Comforti, Y.C. Chung, M. Heiblum, V. Umansky, and D. Mahalu, 
Nature {\bf 416}, 515 (2002).
\bibitem{b29} J.K. Jain, Phys.\ Rev.\ Lett. {\bf 63}, 199 (1989).
\bibitem{b14} R. Guyon, P. Devillard, T. Martin, and I. Safi, 
Phys.\ Rev.\ B {\bf 65}, 153304 (2002).
\bibitem{rem} Note that Eq.~(\ref{e11}) defines only the commutator
of the fields $\eta_j$. Because of the vanishing energy of these
fields, however, the uncertainty in their definition affects only
the zero-frequency component of the imaginary-time correlator
(\ref{e13}) which is not obtained by the analytical continuation
from real time, and would not change Eq.~(\ref{e16}) on which the
subsequent calculations are based.
\bibitem{b15} A. Schmid, Phys.\ Rev.\ Lett. {\bf 51}, 1506 (1983).
\bibitem{b18} Ch. Nayak, M.P.A. Fisher, A.W.W. Ludwig,
and H.H. Lin, Phys. Rev. B {\bf 59}, 15694 (1999).
\bibitem{b16} C.L. Kane and M.P.A. Fisher, Phys.\ Rev.\ B
{\bf 46}, 15233 (1992).
\bibitem{b20} V.V. Ponomarenko and N. Nagaosa, Phys.\
Rev.\ B {\bf 60}, 16865 (1999); Solid State Commun. {\bf 110},
321 (1999).


\end{thebibliography}
\end{document}